\documentclass[conference]{IEEEtran}
\IEEEoverridecommandlockouts

\usepackage{cite}
\usepackage{amsmath,amssymb,amsfonts}
\usepackage{algorithmic}
\usepackage{graphicx}
\usepackage{textcomp}
\usepackage{xcolor}
\usepackage{booktabs}
\usepackage{color,soul}
\usepackage{tabularx,hhline}
\usepackage{multirow}
\usepackage[draft]{hyperref}
\usepackage{tikz}
\usepackage{textcomp}
\usepackage{enumitem}

\usepackage[strings]{underscore}
\usepackage[left=1.62cm,right=1.62cm,top=0.7in]{geometry}
\usepackage{float}
\begin{document}

\title{Optimal Placement and Power Supply of Distributed Generation to Minimize Power Losses}

\author{
\IEEEauthorblockN{\textbf{Shijie Pan}\IEEEauthorrefmark{1}, \textbf{Sajjad Maleki}\IEEEauthorrefmark{2}, \textbf{Subhash Lakshminarayana}\IEEEauthorrefmark{2}, \textbf{Charalambos Konstantinou}\IEEEauthorrefmark{1}}
\IEEEauthorblockA{\IEEEauthorrefmark{1}CEMSE Division, King Abdullah University of Science and Technology (KAUST)\\
\IEEEauthorrefmark{2}School of Engineering, University of Warwick, Coventry, United Kingdom 
}
\IEEEauthorblockA{{
E-mail: \{shijie.pan, charalambos.konstantinou\}@kaust.edu.sa},\\
\{sajjad.maleki-kahnamooyi, subhash.lakshminarayana\}@warwick.ac.uk 
}}

\IEEEaftertitletext{\vspace{-0.8\baselineskip}}
\maketitle
\begin{abstract}
An increasing number of renewable energy-based distribution generation (DG) units are being deployed in electric distribution systems. Therefore, it is of paramount importance to optimize the installation locations as well as the power supply of these DGs. The placement of DGs in the grid can decrease the total distance that power is transmitted and thus reduce power losses. Additionally, the reactive power supply from the DGs can further reduce power losses in the distribution grid and improve power transmission efficiency. This paper presents a two-stage optimization strategy to minimize power losses. In the first stage, the DG locations and active power supply that minimize the power losses are determined. The second optimization stage identifies the optimal reactive power output of the DGs according to different load demands. The proposed approach is tested on the IEEE 15-bus and the IEEE 33-bus systems using DIgSILENT PowerFactory. The results show that the optimized power losses can be reduced from $58.77$ kW to $3.6$ kW in the 15-bus system, and from $179.46$ kW to around $5$ kW in the 33-bus system. Moreover, with the proposed optimization strategy, voltage profiles can be maintained at nominal values  enabling the distribution grid to support higher load demand.
\end{abstract}


\vspace{-1mm}
\section{Introduction}\label{s:introduction}
\vspace{-0.5mm}
When power is transmitted from generation to end users, power losses occur. It is estimated that annual power losses in transmission and distribution (T\&D) systems globally average about 8.12\% of the transmitted power \cite{world2015}.  
Power losses carry significance due to their environmental and economic implications, encompassing costs related to carbon emissions and affecting generation capacity. Moreover, users of the network bear the responsibility of covering these expenses. Distributors are motivated by economic incentives to tackle this issue, as reducing losses leads to additional permitted revenue \cite{sultana2016review}. For example, New York state lost more than 6.6M MWh of electricity in T\&D in 2021, which amounts to a bill of about 1.1b USD worth \cite{NewYork2021}.  
From a system perspective, minimizing power losses decreases the current flow required to be transmitted on the lines, which reduces thermal effects and allows the distribution system to support more load demand. 

In order to reduce the power losses in the distribution grid, existing work considers different perspectives. A number of studies focus on the power supply. For instance, in \cite{qiu1987new,mamandur1981optimal,badar2012reactive,alkaabi2018short}, the authors focus on minimizing the power losses by adjusting the reactive power dispatch. Other works consider the impact of electric vehicles on the distribution grid, e.g., the works in \cite{deilami2011real,luo2014real,singh2010influence}, schedule the charging of electric vehicles to achieve power loss reduction and improve voltage profiles. From the grid topology perspective, works focusing on network reconfiguration aim to reduce power losses by altering the status of switches in the distribution grid \cite{baran1989network,rao2012power,pegado2019radial}. 

In addition, the location of new renewable-based distribution generation (DG) units in the grid could also be optimized to minimize power losses, considering also the size of the renewable energy market. According to \cite{irena2023renewable}, global renewable capacity increased by $295$ GW in 2022. This number will continue to grow in the following years. 
Power will flow over additional lines in a distribution grid with poorly placed generations, e.g., at the end of a long branch, leading to additional losses. Therefore, it is essential to place DGs in optimal locations to minimize power losses. In \cite{lalitha2010optimal}, the authors use a two-stage optimization, which first identifies the optimal locations for DGs and then determines the size of those DGs. The optimal placement of DGs considering different types of DG sources is considered in \cite{kansal2013optimal}. An approach from the DG owners' perspective is presented in \cite{avar2021optimal}. The authors aim to find the optimal DG sizes and locations that maximize DG owners' economic benefits. In addition to economic aspects, the environmental issues are also considered in \cite{melgar2018adaptive} to promote low carbon emission systems.

Power losses can be reduced by optimizing the placement of DGs. Also, reactive power adjustment can achieve further power loss reduction during system operation. 
However, limited research has been undertaken that examines how to utilize both methods. In this paper, we propose a two-stage mixed-integer second-order cone programming (MISOCP) optimization problem to reduce power losses in distribution systems. First, the problem identifies the optimal placement for renewable-based DG installations as well as their active power outputs. Then, it optimizes the reactive power supply from those DGs to match the load demand. We examine the effects of the two stages on reducing power losses in the grid sequentially. We also consider, following the study in \cite{van2016linear}, conservative voltage reduction (CVR) to adjust the loads' voltages to reduce power losses. Furthermore, we utilize the ZIP load model to capture the relationship between voltage and power demands in CVR. The approach is verified using two IEEE distribution benchmarks in DIgSILENT PowerFactory, in which PVs are the considered renewable-based DG units. The results show that power losses in both systems are significantly reduced with the proposed optimization approach. Moreover, voltage profiles are maintained at nominal values. 
 

The rest of this paper is organized as follows. Section \ref{s:related-work} introduces preliminaries, and Section \ref{s:methodology} presents the details of the formulated optimization. Section \ref{s:results} demonstrates the simulation results, while Section \ref{s:conclusion} concludes the paper.

\vspace{-2mm}
\section{Preliminaries}\label{s:related-work}
\vspace{-1mm}
\subsection{Conservation Voltage Reduction (CVR)}\label{ss:cvr}
\vspace{-0.5mm}

CVR is a cost-effective way to save energy {\cite{wang2013review}}. It can reduce peak demand and power losses as well as achieve more energy savings by lowering voltages in the distribution grid. According to the American National Standards Institute (ANSI), the voltage range at the distribution transformer secondary terminals is set to $120$ V $\pm5$\% \cite{ANSI1995ANSI}. Within the lower half of the voltage band, i.e., $-5$\% to $0$\%, voltages can be adjusted cost-effectively and will not cause damage to consumer appliances \cite{kennedy1991conservation, wang2013review}. Furthermore, CVR-enabled appliances are usually designed to consume less energy when operating at lower voltages. It usually ranges from $0.3$\% to $1$\% load reduction per $1$\% voltage reduction \cite{wang2013review}. Thus, reducing the voltages of the distribution grid through CVR can reduce load demand and power consumption.

\vspace{-2mm}
\subsection{ZIP Load Model}\label{ss:zip}
\vspace{-0.5mm}
The ZIP load model is a static model that represents the voltage dependency of the power consumed by a load. It is comprised of constant impedance \textit{Z}, constant current \textit{I}, and constant power \textit{P}, which are also called the \textit{Z}, \textit{I}, \textit{P} coefficients, respectively. In the experiments of \cite{rahimi2012evaluation}, the P-V and Q-V relationships of the most commonly used appliances are tested under varying voltage conditions. The authors demonstrate that all loads have variability with different voltage levels based on the \textit{Z}, \textit{I}, \textit{P} coefficients\cite{hossan2017comparison}:
\vspace{-2mm}
\begin{equation}\label{eq1}
\left\{
    \begin{aligned}
    &P^{ZIP} = P_0(Z_P\cdot {V}^2+I_P\cdot {V}+P_P),\\
    &Q^{ZIP} = Q_0(Z_Q\cdot {V}^2+I_Q\cdot {V}+P_Q)
    \end{aligned}
\right.
\vspace{-2mm}
\end{equation}where $P^{ZIP}$ and $Q^{ZIP}$ denote the active and reactive power modeled by the ZIP model, respectively. $P_0$, $Q_0$ are the base values, $Z_P$, $I_P$, $P_P$, $Z_Q$, $I_Q$, $P_Q$ are the ZIP coefficients of the active power and reactive power, respectively, and $Z_P+I_P+P_P=1$, $Z_Q+I_Q+P_Q=1$. $V$ is the measured voltage, and also the variable of the model.

\vspace{-1.5mm}
\section{Methodology}\label{s:methodology}
\begin{figure}[t]
    \centering
    \includegraphics[width=0.6\linewidth]{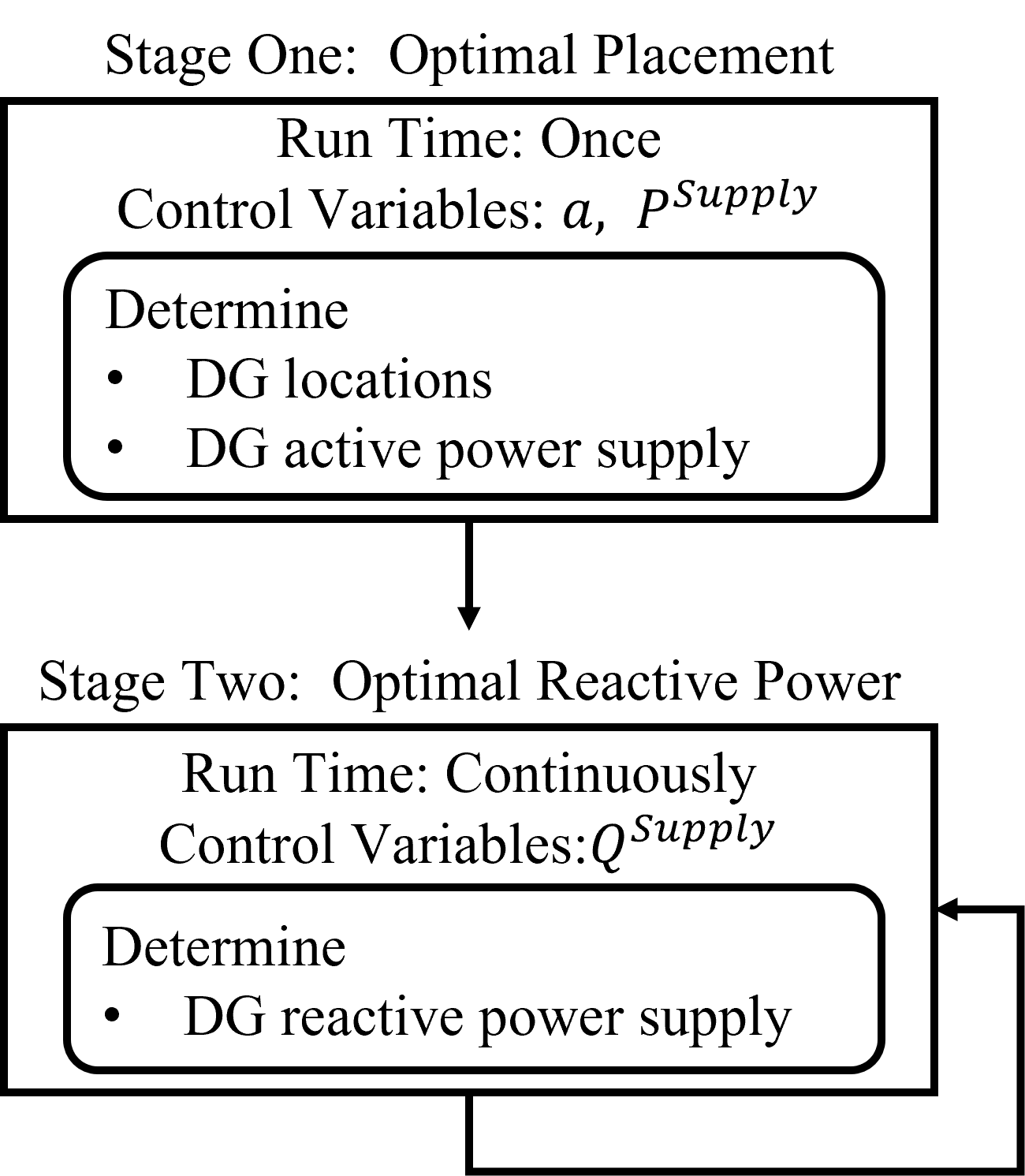}
    \vspace{-2mm}
    \caption{Schematic diagram of the proposed two-stage optimization for optimal DG placement and power supply to minimize power lossses.}
    \vspace{-4mm}
    \label{fig:framework}
\end{figure}

\vspace{-0.8mm}
In this section, we present the two-stage optimization to minimize the power losses of the DG-based distribution grid. The framework of the optimization is depicted in Fig.~\ref{fig:framework}. 

\vspace{-2mm}
\subsection{Stage 1: Optimal DG Placement}\label{ss:locations}
\vspace{-0.5mm}
The placement of renewable-based DGs in the distribution grid is significant since it can affect the transmission path that power is delivered to the end users. By optimizing the placement of DGs, power can flow through ``shorter paths'', thus, the power losses can be minimized. {Let us consider a boolean variable {$a \in \{0,1\}^{N_{Bus}}$}, where $N_{Bus}$ is the number of buses, and $a_i$ denotes whether or not there exists a renewable-based DG at bus $i$. In addition, $P^{{Supply}_i} \in \mathbb{R}^{N_{Bus}}$ represents the active power supply at bus $i$. } The objective of the optimization is to determine the placement of DGs at the buses and their active power output, and can be expressed as follows:
\vspace{-2mm}
\begin{equation}\label{eq2}
\vspace{-1mm}
\mathop {\min }\limits_{\textbf{$a$,\ $P^{Supply}$}}
\sum_{i,j \in N_{Bus}} {I^2_{ij} r_{ij}}
\end{equation} where $I_{ij}$ and $r_{ij}$ are the current transmitted and resistance on the line connecting bus $i$ and bus $j$, respectively. In the paragraphs below, we present the details of the various constraints which such optimization model need to satisfy.

Eq.~\eqref{eq3} bounds the active and reactive power  output of the DGs. {$\overline{S^{DG}}$  refers to the maximum apparent power that the DG can continuously supply. It is relevant to the type of DG and should be determined before the optimization.} Eq.~\eqref{eq4} shows that there are $N_{DG}$ renewable-based DGs in the system.    
\vspace{-1mm}
\begin{equation}\label{eq3}
    {P^{Supply}_i}^2+{Q^{Supply}_i}^2 \le a_i\cdot \overline{S^{DG}}^2 
\end{equation}
\begin{equation}\label{eq4}
\vspace{-2mm}
    \sum_{i \in N_{Bus}}a_{i} = N_{DG}
\end{equation}

The power flow model used in our work is the DistFlow model, a branch flow framework developed to help analyze the power flow in a radial grid \cite{baran1989network}. Eq.~\eqref{eq5} represents the branch flow equations for both active and reactive power. $P_{ij}$ and $Q_{ij}$ are the active and reactive power transmitted between bus $i$ and bus $j$. $P^{Load}_j$, $P^{Supply}_j$ are the load demand and power supply at bus $j$, similar to $Q^{Load}_j$ and $Q^{Supply}_j$.
\vspace{-1mm}
\begin{equation}
\vspace{-1mm}\label{eq5}
\left\{
    \begin{aligned}
    &P_{ij} = P^{Load}_j - P^{Supply}_j + r_{ij}I^2_{ij} + P_{jk},\\
    &Q_{ij} = Q^{Load}_j - Q^{Supply}_j + x_{ij}I^2_{ij} + Q_{jk}
    \end{aligned}
\right.
\end{equation}The loads in the simplified DistFlow model are considered to have constant power. However, as presented in \ref{ss:zip}, CVR-enabled devices have different load demands depending on their operating voltages. Therefore, instead of the constant-power loads, we use the ZIP load model to better characterize the relationship between load demand and voltage. 
In order to find a global minimum for the optimization, the ZIP load model is relaxed into a convex form. 
We follow the strategy in \cite{nazir2020approximate} to approximate the ZIP model with an equivalent linearized ZP model (Eq. \ref{eq6}). It is based on the principle that the bus voltages are relatively close to one per unit (p.u.). The derivation of the active power demand term is given as follows:
\vspace{-1mm}
{\begin{align}
\vspace{-4mm}
    P^{ZIP}_j &= P^{Load}_j(Z_P\cdot V_j^2+I_P\cdot V_j+P_P)\notag\\
    &\approx P^{Load}_j(Z_P\cdot V_j^2+I_P\cdot (1+\frac{V_j^2-1}{2})+P_P)\notag\\
    &\approx P^{Load}_j((Z_P+\frac{I_P}{2})V_j^2+(P_P+\frac{I_P}{2}))\label{eq6}
\end{align}}
Considering also the reactive power, we may rewrite Eq.~\eqref{eq5} as follows:
\vspace{-1mm}
\begin{equation}
\vspace{-1mm}
\label{eq7}
\left\{
    \begin{aligned}
    &P_{ij} = P^{ZP}_j - P^{Supply}_j + r_{ij}I^2_{ij} + P_{jk},\\
    &P^{ZP}_j = P^{Load}_j((Z_P+\frac{I_P}{2})V_j^2+(P_P+\frac{I_P}{2})), \\
    &Q_{ij} = Q^{ZP}_j - Q^{Supply}_j + x_{ij}I^2_{ij} + Q_{jk}, \\
    &Q^{ZP}_j = Q^{Load}_j((Z_Q+\frac{I_Q}{2})V_j^2+(P_Q+\frac{I_Q}{2}))
    \end{aligned}
\right.
\end{equation}

Moreover, we consider the correlation of voltages between two buses. $V_i$ and $V_j$ are the voltages at bus $i$ and bus $j$, and $r_{ij}$ and $x_{ij}$ are the resistance and reactance between these two buses, respectively. This is represented as follows: 
\vspace{-1mm}
\begin{equation}
\vspace{-1mm}\label{eq8}
    V^2_j = V^2_i + (r_{ij}^2 + x_{ij}^2)I^2_{ij} - 2(r_{ij}P_{ij} + x_{ij}Q_{ij})
\end{equation}


Similar to the ZIP model, the complex power flow relationship between bus $i$ and bus $j$, $S_{ij} = V_iI^*_{ij}$, also needs to be restricted as convex, represented as follows: 
\vspace{-1mm}
\begin{align} 
\vspace{-2mm}
    S_{ij} = V_iI^*_{ij} 
    \Longrightarrow &S_{ij}S_{ij}^\ast=V_iV_i^\ast I_{ij}I_{ij}^\ast \notag\\  
    \Longrightarrow &\left(P_{ij}+jQ_{ij}\right)\left(P_{ij}-jQ_{ij}\right)=\left|V_i\right|^2\left|I_{ij}\right|^2 \notag\\
    \Longrightarrow &P_{ij}^2+Q_{ij}^2={V^2_i}I^2_{ij} \notag\\
    \Longrightarrow &I^2_{ij} \geq \frac{P^2_{ij}+Q^2_{ij}}{V^2_i}\label{eq10}
\end{align}Eq.~\eqref{eq10} can be further derived into a second-order cone programming (SOCP) form, thus formulating convex relaxations of the optimization problem~\cite{sun2022multi}.
\vspace{-1mm}
\begin{align}
\vspace{-5mm} 
    I^2_{ij} \geq \frac{P^2_{ij}+Q^2_{ij}}{V^2_i} 
    \Longrightarrow &P_{ij}^2+Q_{ij}^2\le\frac{\left(v_i+l_{ij}\right)^2-\left(v_i-l_{ij}\right)^2}{4}\notag\\
    \Longrightarrow
    &\left\lVert \begin{aligned}
    2&P_{ij}\\
    2&Q_{ij}\\
    I^2_{ij}&-V^2_i
    \end{aligned} \right\rVert \le I^2_{ij}+V^2_i\label{eq11}
\end{align}

The current and voltage limits are also considered, as shown in Eq.~\eqref{eq12} and Eq.~\eqref{eq13}:
\vspace{-2mm} 
\begin{equation}\label{eq12}
\vspace{-2mm} 
    I^2_{ij} \le \overline{I_{ij}}^2
\end{equation}
\begin{equation}\label{eq13}
\vspace{-3mm} 
    \underline{V_{i}}^2 \le V^2_{i} \le \overline{V_{i}}^2
\end{equation}where  $\overline{I_{ij}}$ is the maximum current that can be transmitted on the line connecting bus $i$ and bus $j$. $\underline{V_{i}}$ and $\overline{V_{i}}$ are the upper and lower limit of the voltage at bus $i$.

With the objective function and relaxed constraints above, the optimization problem can be written as follows: 
\vspace{-1mm} 
\begin{equation}\label{eq14}
\vspace{-1mm} 
\begin{aligned}
&\mathop {\min }\limits_{ \textbf{$a$,\ $P^{Supply}$}}
\sum_{i,j \in N_{Bus}} {I^2_{ij} r_{ij}}\\
\textrm{s.t.} \qquad&(\ref{eq3})(\ref{eq4})(\ref{eq7})(\ref{eq8})(\ref{eq11})(\ref{eq12})(\ref{eq13})
\end{aligned}
\end{equation} The MISOCP problem in Eq.~\eqref{eq14} aims to determine the optimal placement and active power supply of renewable-DGs in order to minimize the power losses in the distribution grid. 

\vspace{-2mm} 
\subsection{Stage 2: Optimal Reactive Power Output}\label{ss:reactive}
\vspace{-0.5mm} 

The previous stage focuses on the placement and sizing of the DGs that should be determined before the distribution grid operation. In the second stage, a reactive power control strategy is applied. The optimal reactive power of the DGs is updated periodically to follow the different load profiles. This ensures that the reactive power supply meets the demand and reduces the power losses. The stage 2 optimization is shown below:  
\vspace{-0.5mm} 
\begin{equation}\label{eq15}
\vspace{-2mm} 
\begin{aligned}
&\mathop {\min }\limits_{ \textbf{$Q^{Supply}$}}
\sum_{i,j \in N_{Bus}} {I^2_{ij} r_{ij}}\\
\textrm{s.t.} \qquad&(\ref{eq3})(\ref{eq4})(\ref{eq7})(\ref{eq8})(\ref{eq11})(\ref{eq12})(\ref{eq13})
\end{aligned}
\end{equation}The control variable in this stage is {$Q^{Supply} \in \mathbb{R}^{N_{Bus}}$, the set of the reactive power being supplied at each bus. Similar to $P^{Supply}$, if there is no DG at bus $i$, $Q^{Supply}_i=0$.} The objective and constraints remain the same as in stage 1.

\vspace{-0.5mm}
\section{Experimental Results}\label{s:results}
\vspace{-0.5mm}

\begin{figure}[!t]
    \centering
    \includegraphics[width=0.85\linewidth]{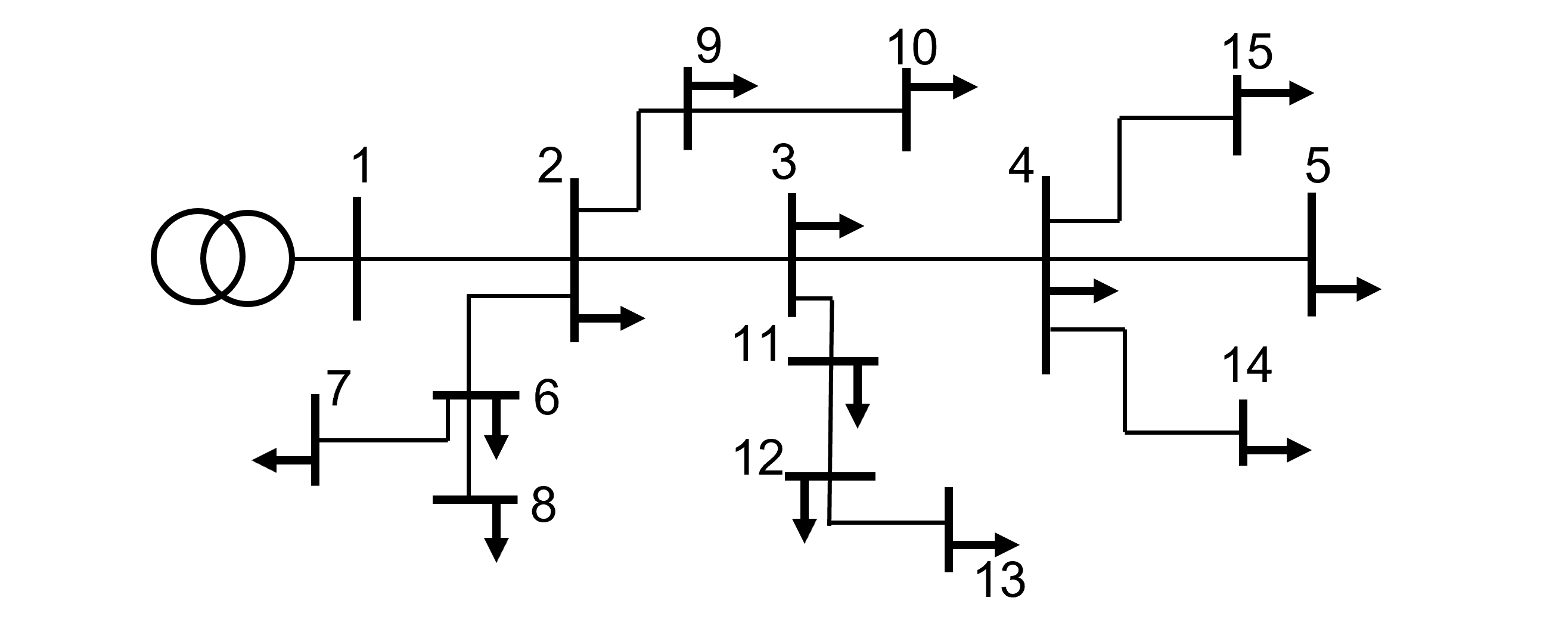}
    \vspace{-4mm}
    \caption{Single-line diagram of the IEEE 15-bus distribution system.}
    \vspace{-2mm}
    \label{fig:IEEE15}
\end{figure}

\begin{figure}[!t]
    \centering
    \includegraphics[width=0.95\linewidth]{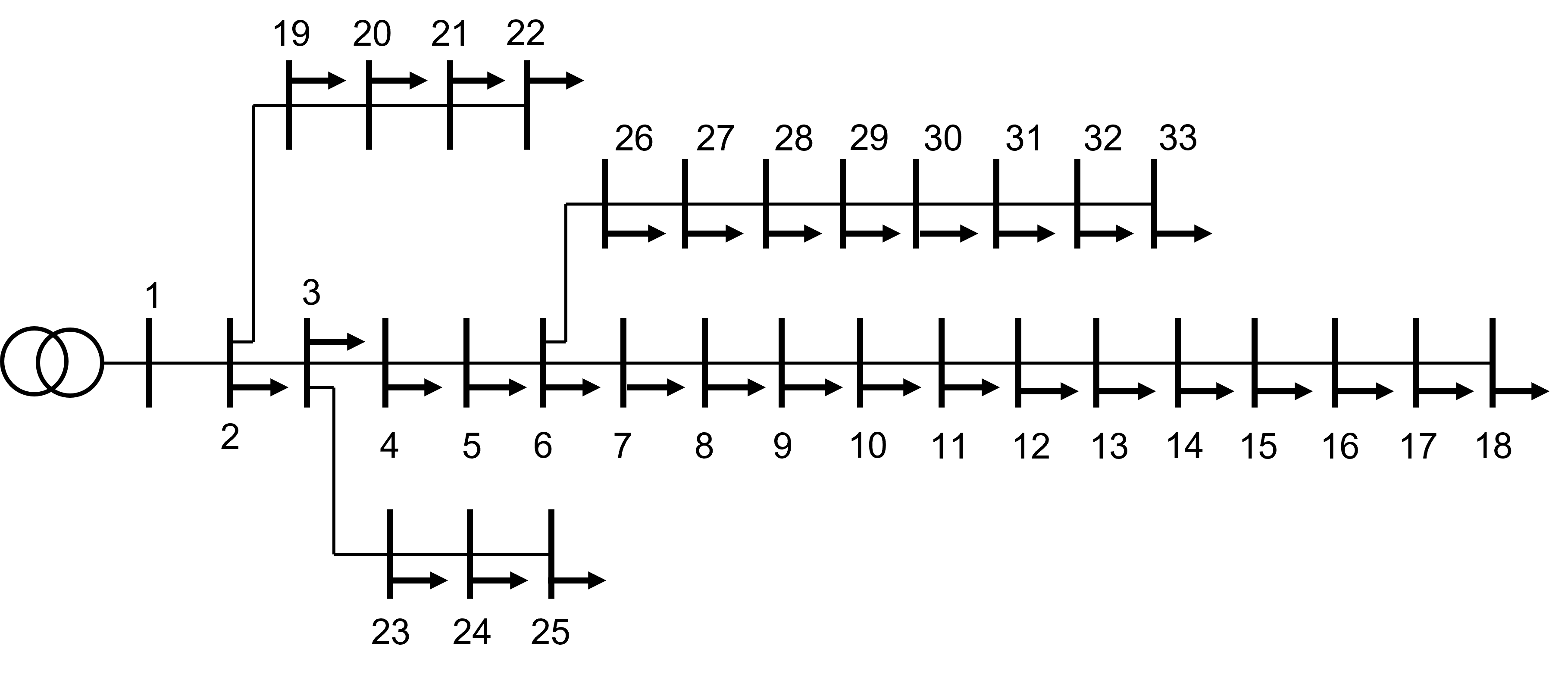}
    \vspace{-4mm}
    \caption{Single-line diagram of the IEEE 33-bus distribution system.}
    \vspace{-4mm}
    \label{fig:IEEE33}
\end{figure}

In this section, we test the proposed two-stage optimization problem in two radial distribution grids. The renewable-based DGs in both benchmarks are simulated using PV units. 
The topology of the IEEE 15-bus and 33-bus distribution systems are shown in Fig.~\ref{fig:IEEE15} and Fig.~\ref{fig:IEEE33}, respectively. For the IEEE 15-bus system, the base voltage is $11$ kV, and the base power is $100$ {kVA}. The line parameters and load data are retrieved from \cite{sudhakar2016modeling}. The proposed optimization is programmed in Python and solved by CPLEX. We then feed the optimal placement and reactive power outputs of the DGs into DIgSILENT PowerFactory using its supported Python interface. The power losses and other experimental data are calculated by DIgSILENT PowerFactory. The IEEE 33-bus distribution system is also used to validate the proposed method in a larger-scale system. 
The base voltage for the IEEE 33-bus system is $12.66$ kV, and the base power is $10$ MVA~\cite{dolatabadi2020enhanced}. 

\begin{figure}[t]
    \centering
    \includegraphics[width=\linewidth]{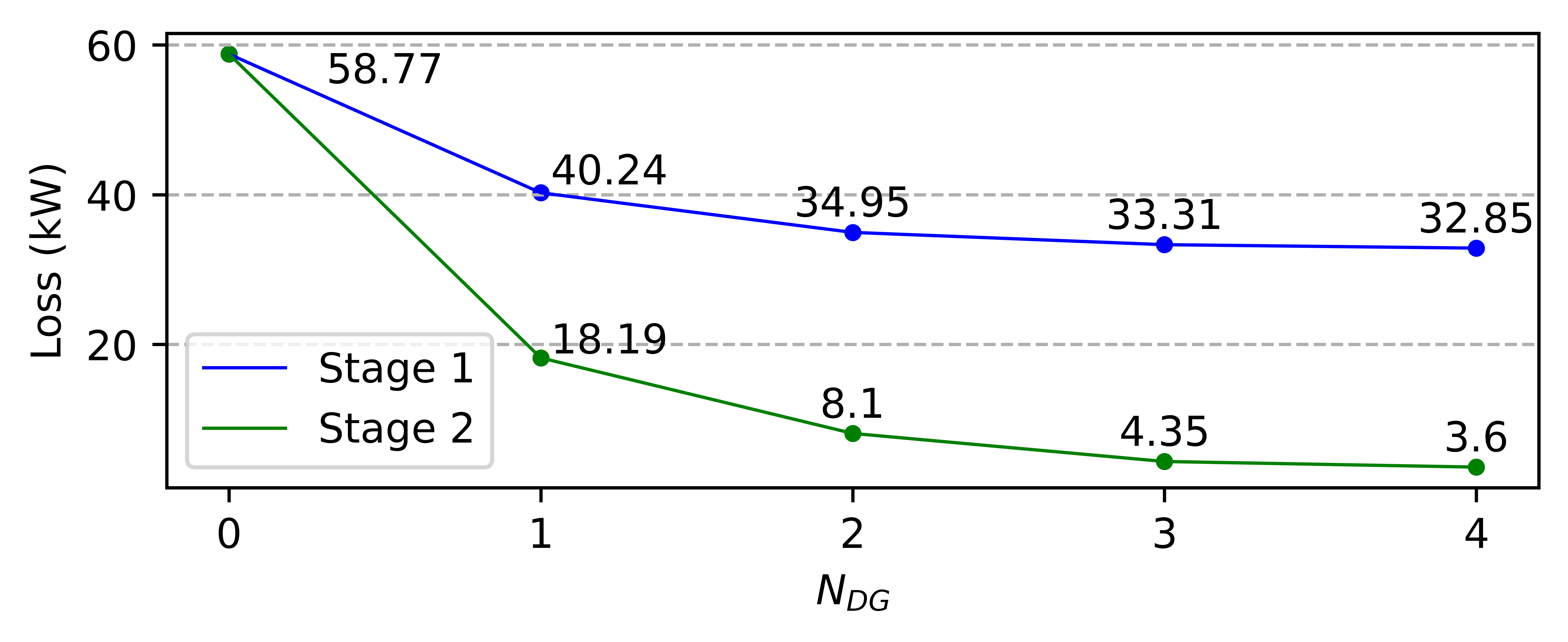}
    \vspace{-8mm}
    \caption{The power flow losses with different $N_{DG}$ in the IEEE 15-bus system after different optimization stages. Stage 1: With optimal locations and active power outputs. Stage 2: With additional optimal reactive power outputs. $N_{DG}=0$ can be regarded as without any optimization.}
    \vspace{-3mm}
    \label{fig:Loss15}
\end{figure}

\begin{figure}[t]
    \centering
    \includegraphics[width=\linewidth]{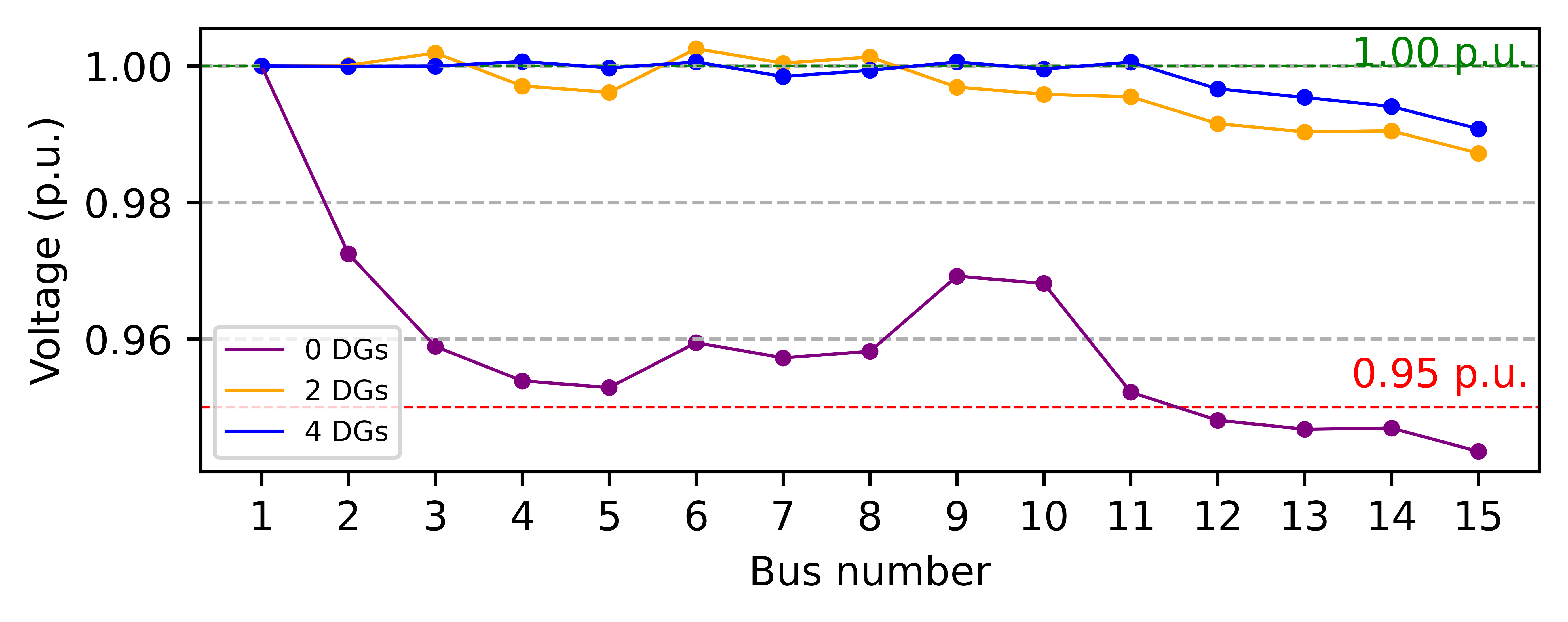}
    \vspace{-8mm}
    \caption{The voltage profiles of each bus with $N_{DG}=0, 2, 4$ incorporated in the IEEE 15-bus system after stage 2 optimization.}
    \vspace{-5mm}
    \label{fig:Voltage15}
\end{figure}

\vspace{-2mm} 
\subsection{Case Studies}\label{ss:base}
\vspace{-0.5mm} 
The effect of each stage of the proposed optimization is evaluated by applying our approach to the test systems. We start without incorporating any DG in the benchmarks. The sole supply of the power is from the transmission system that is connected to bus 1. Power is transmitted over a long distance before it reaches the end users, which increases the power losses. This case provides the power losses and voltage profiles of the system without any optimization. Then, we perform experiments with only the first stage of the proposed optimization in order to study how the placement of DGs affects the power losses by identifying the DGs' optimal locations and active power outputs. At last, we apply the second stage of the optimization to determine the optimal reactive power of the DGs based on different load demands. 

\begin{figure}[!t]
    \centering
    \includegraphics[width=\linewidth]{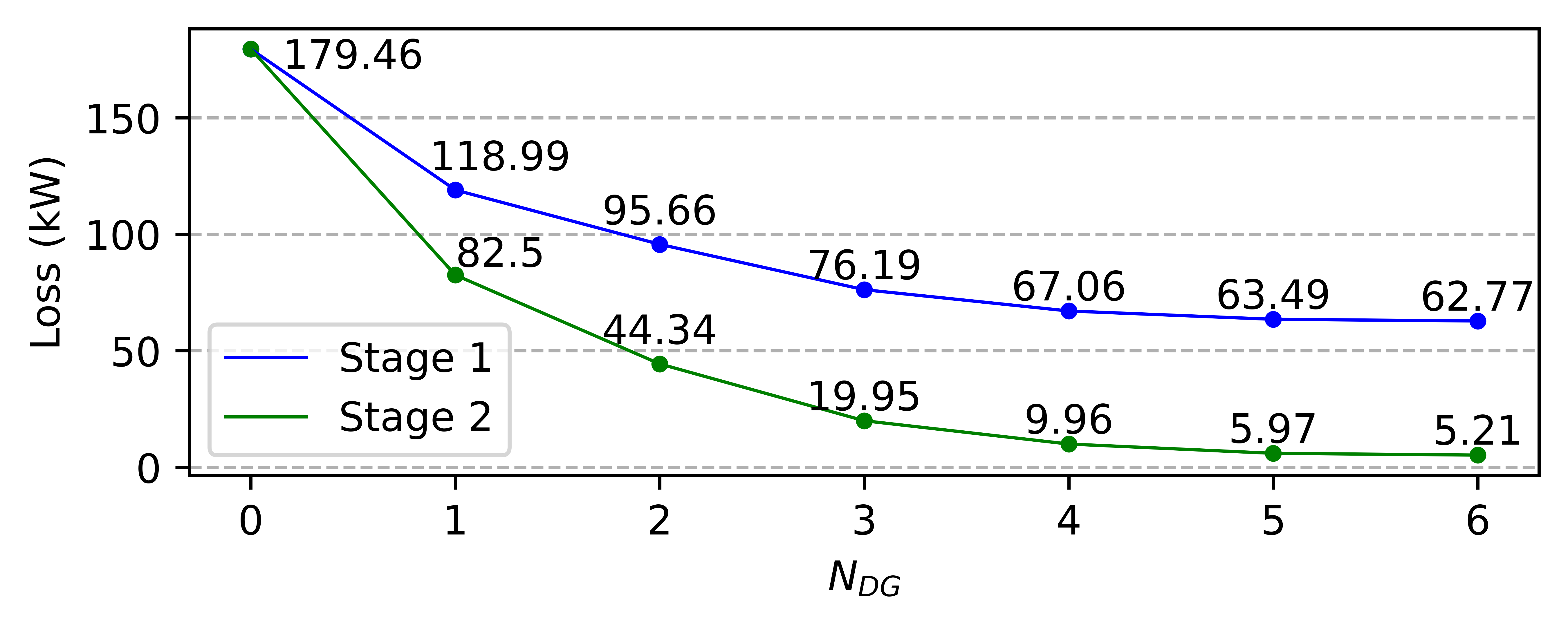}
    \vspace{-8mm}
    \caption{The power flow losses with different $N_{DG}$ in IEEE 33-bus system after different optimization stages. Stage 1: With optimal locations and active power outputs. Stage 2: With additional optimal reactive power outputs. $N_{DG}=0$ can be regarded as without any optimization.}
    \vspace{-3mm}
    \label{fig:Loss33}
\end{figure}

\begin{figure}[!t]
    \centering
    \includegraphics[width=\linewidth]{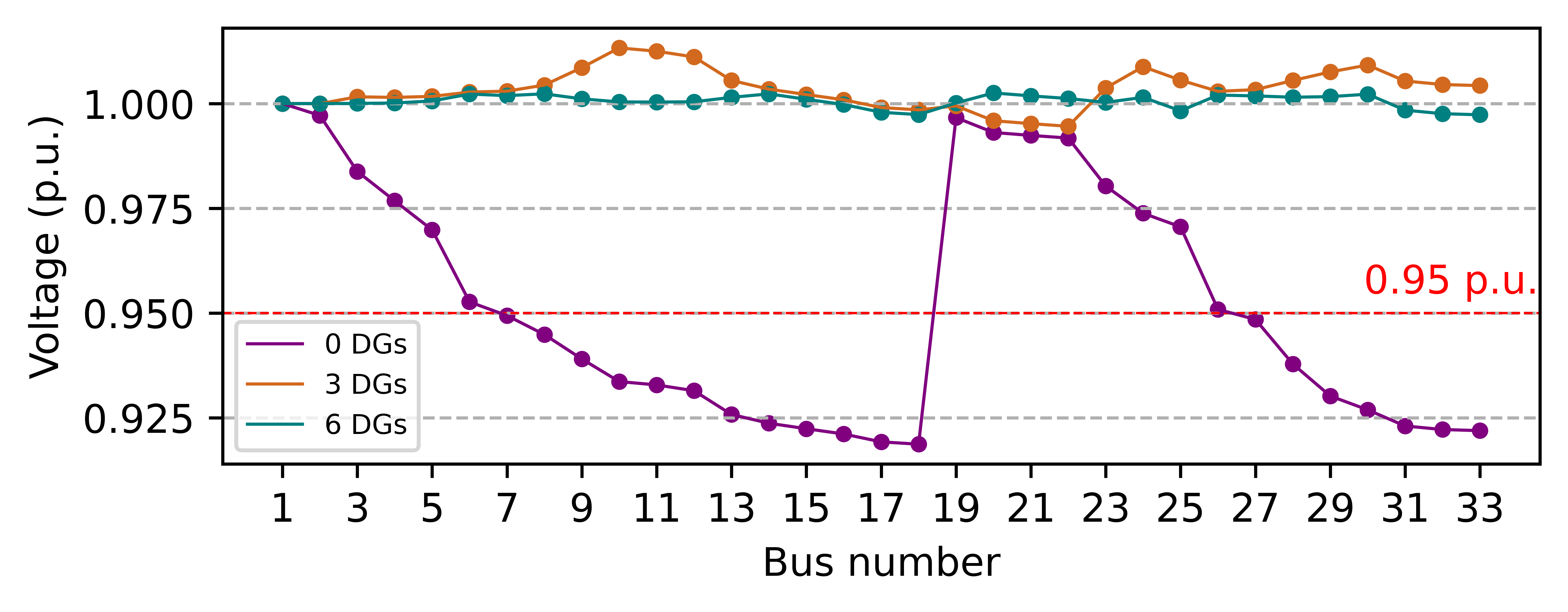}
    \vspace{-8mm}
    \caption{The voltage profiles of each bus with $N_{DG}=0, 3, 6$ incorporated in the IEEE 33-bus system after stage 2 optimization.}
    \vspace{-5mm}
    \label{fig:voltage33}
\end{figure}

\subsubsection{Base Case}\label{sss:base}

In this case, we calculate the power losses without any DG in the distribution system, while also considering CVR. For the IEEE 15-bus system, it is clear from Fig.~\ref{fig:Loss15} that {the power losses in this base case are $58.77$ kW (the highest value in the figure)}. The {voltage} profiles in this base case are presented in Fig.~\ref{fig:Voltage15}. It can be observed (purple line) that the voltages drop significantly and violate the limit ($0.95$ p.u.) from buses 12 to 15. 
As for the IEEE 33-bus system, power losses reach $179.46$ kW without any DG in the system (Fig.~\ref{fig:Loss33}). 
Fig.~{\ref{fig:voltage33}} shows the voltage profiles of this benchmark case which drop significantly as moving farther away from bus 1. {The voltages at buses 7-18 and 27-33 are below $0.95$ p.u.}

\begin{table}[t]
    \centering  
    \caption{{Optimal locations and active power outputs with different $N_{DG}$ incorporated in the IEEE 15-bus system.}}  
    \label{table1}  
    \begin{tabular}{cccc}  
    \hline  
		$N_{DG}$& Locations (Bus)& Active Power Outputs (100kW) \\ 
    \hline
        0&/&/\\
        1&3&11.4\\
        2&3, 6&7.2, 4.2\\
        3&4, 6, 11&3.9, 4.4, 2.8\\
        4&4, 6, 9, 11&3.4, 3.7, 1.4, 2.8\\
    \hline
    \end{tabular}
    \vspace{-3mm}
\end{table}

\subsubsection{Stage 1}\label{sss:stage1}

In the first step of the optimization, we aim to find the optimal locations and the active power outputs of the DGs. We test our approach with different numbers of DGs in the grid. Table \ref{table1} lists the locations {and active power outputs} to install the DGs optimally. As for the power losses, it can be observed from Fig.~\ref{fig:Loss15} that, even with only one DG, the power losses can be reduced by 1/3 compared to the base case. This is mainly due to the DG installed at bus 3, which helps in decreasing the total transmission distance of the power and reduces power losses. The power losses decrease gradually as more DGs are incorporated, which balances the power supply across the distribution grid. However, it is clear from Fig.~\ref{fig:Loss15} that the reduction in power losses slows down as the number of DGs increases and becomes insignificant (less than $1$ kW) with 4 DGs in the system. {Installing new DGs beyond this number does not necessarily reduce the power losses. Further reduction requires other methods, e.g., adjusting the reactive power outputs of the DGs.} 
Similarly, for the IEEE 33-bus system, the optimal locations {and active power outputs} of DGs are listed in Table \ref{table2}. {Fig.~\ref{fig:Loss33}} plots the power losses with different numbers of DGs being incorporated into the system. The blue line in {Fig.~\ref{fig:Loss33}} indicates how the losses are gradually reduced from $179.46$ kW to $62.77$ kW by installing new DGs with the optimal placement approach. Although the losses significantly reduce as the number of DGs in the system increases, the reduction reaches a plateau with 6 DGs installed in the system, i.e., the power losses are only reduced by $0.72$ kW when the sixth DG is connected to the grid.

\subsubsection{Stage 2}\label{sss:stage2} 
In the first stage of the optimization, the reactive power supply has not been considered. Here, we further optimize the DGs' reactive power outputs to match the load demands. As shown in Fig.~ \ref{fig:Loss15}, with the optimal reactive power supply from DGs, the power losses in IEEE 15-bus system can be further reduced to less than $4$ kW. This is due to the fact that the optimization balances the power demand and supply. We further study the voltage profiles of the system with 2  and 4 DGs incorporated, as shown in Fig.~\ref{fig:Voltage15}. It is clear that as the number of installed DGs increases, the voltage variations get reduced. 
As for the IEEE 33-bus system, such improvements also exist. Fig.~\ref{fig:Loss33} shows that the power losses are further decreased after adjusting the reactive power outputs of the DGs. In the case with 6 DGs, the power losses are reduced from $179.46$ kW to $62.77$ kW in stage 1, and further reduced from $62.77$ kW to $5.21$ kW in stage 2. The voltage profiles with 3 and 6 DGs installed in the 33-bus system are shown in Fig.~{\ref{fig:voltage33}}. It can be observed that all the bus voltages are above $0.95$ p.u. after incorporating 3 DGs into the system. With 6 DGs installed in the system, the voltage profiles are more closer to the nominal value. 

\begin{table}[t]
    \centering  
    \caption{{Optimal locations and active power outputs with different $N_{DG}$ incorporated in the IEEE 33-bus system.}}  
    \label{table2}  
    \begin{tabular}{cccc}  
    \hline  
		$N_{DG}$& Locations (Bus) & Active Power Outputs (MW) \\ 
    \hline
        0&/&/\\
        1&6&3.723\\
        2&3, 30&2.553, 1.162\\
        3&10, 24, 30&1.171, 1.500, 1.047\\
        4&6, 14, 24, 30&1.103, 0.660, 1.187, 0.767\\
        5&7, 14, 20, 24, 30&0.762, 0.602, 0.485, 1.039, 0.828\\
        6&6, 8, 14, 20, 24, 30&0.548, 0.424, 0.535, 0.446, 0.991, 0.770\\
    \hline
    \end{tabular}
    \vspace{-4mm}
\end{table}

\vspace{-2mm} 
\subsection{Impact of Load Changes}\label{ss:loadIncrease}
\vspace{-1mm} 

\begin{figure}[!t]
    \centering
    \includegraphics[width=\linewidth]{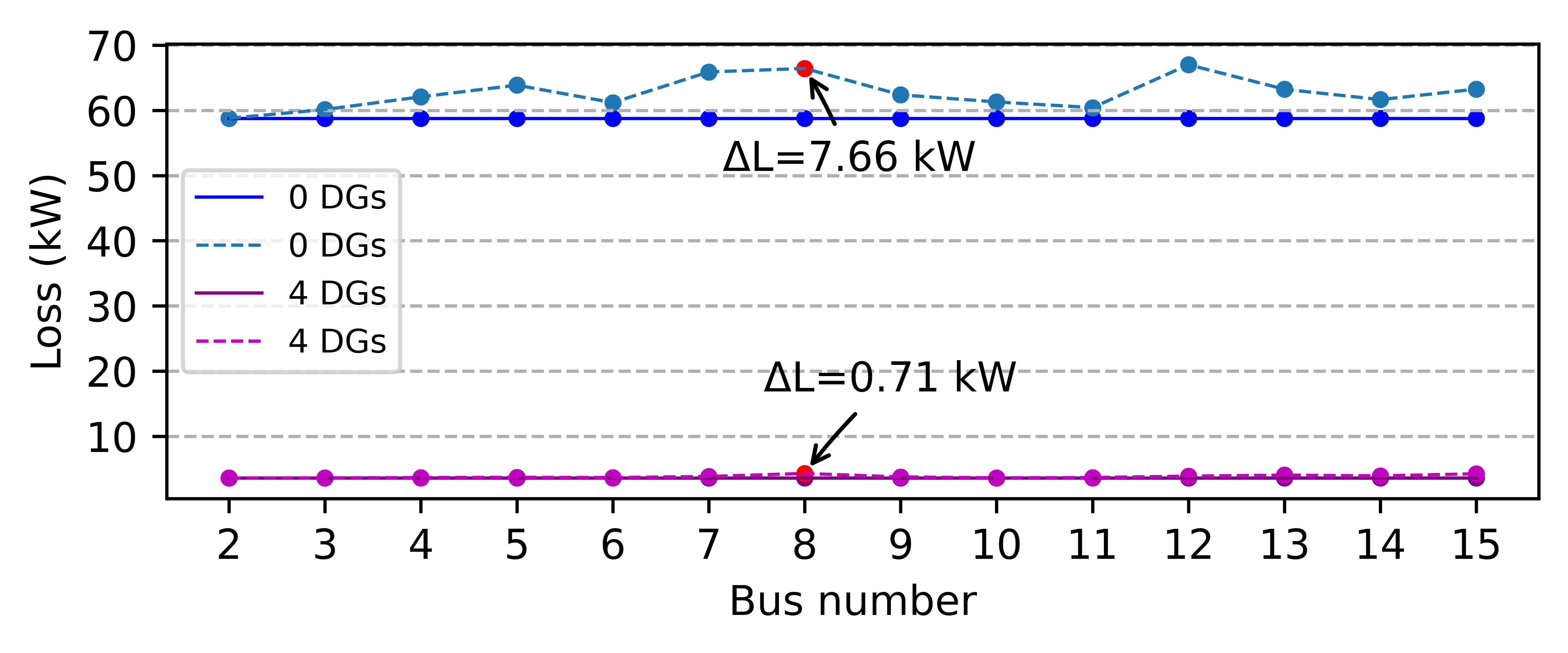}
    \vspace{-7mm}
    \caption{Variations in power losses due to a 50\% load increase at different buses of the IEEE 15-bus system. The solid lines and dash lines represent the power losses before and after the load increase, respectively.}
    \vspace{-2mm}
    \label{fig:load15}
\end{figure}

\begin{figure}[!t]
    \centering
    \includegraphics[width=\linewidth]{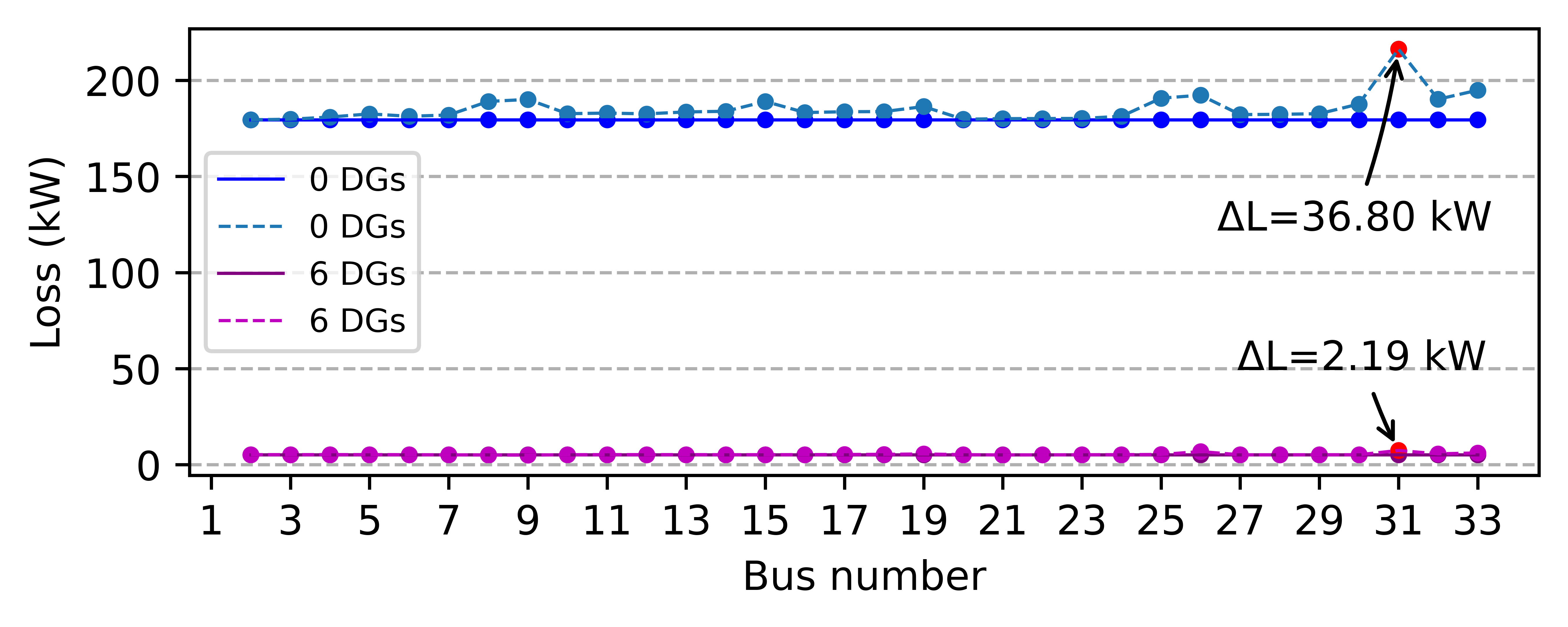}
    \vspace{-7mm}
    \caption{Variations in power losses due to a 50\% load increase at different buses of the IEEE 33-bus system. The solid lines and dash lines represent the power losses before and after the load increase, respectively.}
    \vspace{-4mm}
    \label{fig:load33}
\end{figure}

Load variations can cause circuit overflows and other malfunctions or even damage power system equipment. Therefore, in this part, we consider changes in load demand and study how these affect the distribution grid. Specifically, we sequentially increase the load in each of the load buses of the system by 50\%. For example, for the IEEE 15-bus system, which has 14 loads, we perform the calculation 14 times, one for each of the loads. In each step, we increase the load demand of the bus by 50\%. Then, we obtain the power losses before the optimization is applied and without any DGs in the system, and after the two-stage optimization with the DGs incorporated. We consider as $\Delta L$ the difference in power losses before and after the load increase. Fig.~\ref{fig:load15} shows the variation in power losses for the IEEE 15-bus system. We observe that the increase of load demand at bus 8 leads to $\Delta L=7.66$ kW without any DG in the system. This value is reduced to $0.71$ kW after the optimization. Similarly, for the IEEE 33-bus system, the variations in power losses are shown in Fig.~{\ref{fig:load33}}. With the optimization, the $\Delta L$ at bus 31 is decreased from $36.80$ kW to $2.19$ kW. 

\begin{figure}[ht]
    \centering
    \includegraphics[width=\linewidth]{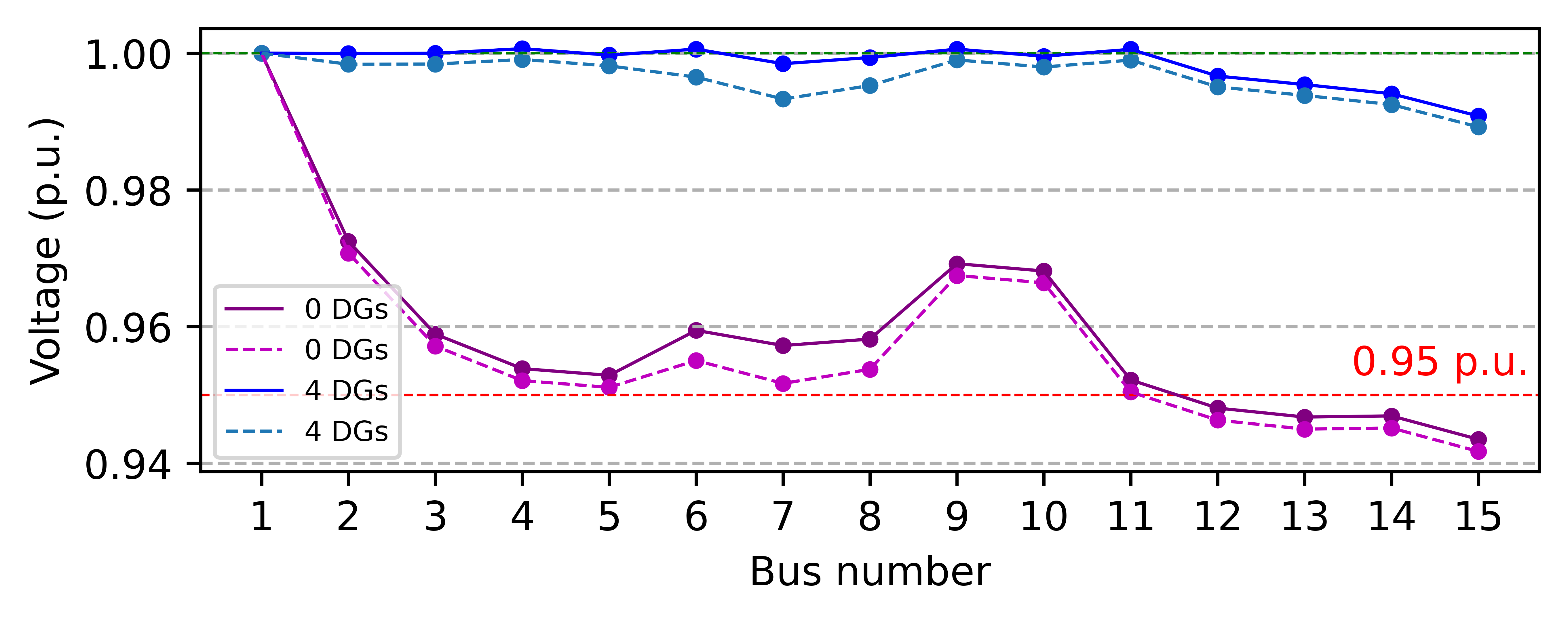}
    \vspace{-7mm}
    \caption{{Bus voltages under a load increase at bus 8 of the IEEE 15-bus system. The solid lines and dash lines represent the voltage magnitudes before and after the load increase, respectively.}}
    \vspace{-2mm}
    \label{fig:voltageLoad15}
\end{figure}

\begin{figure}[ht]
    \centering
    \includegraphics[width=\linewidth]{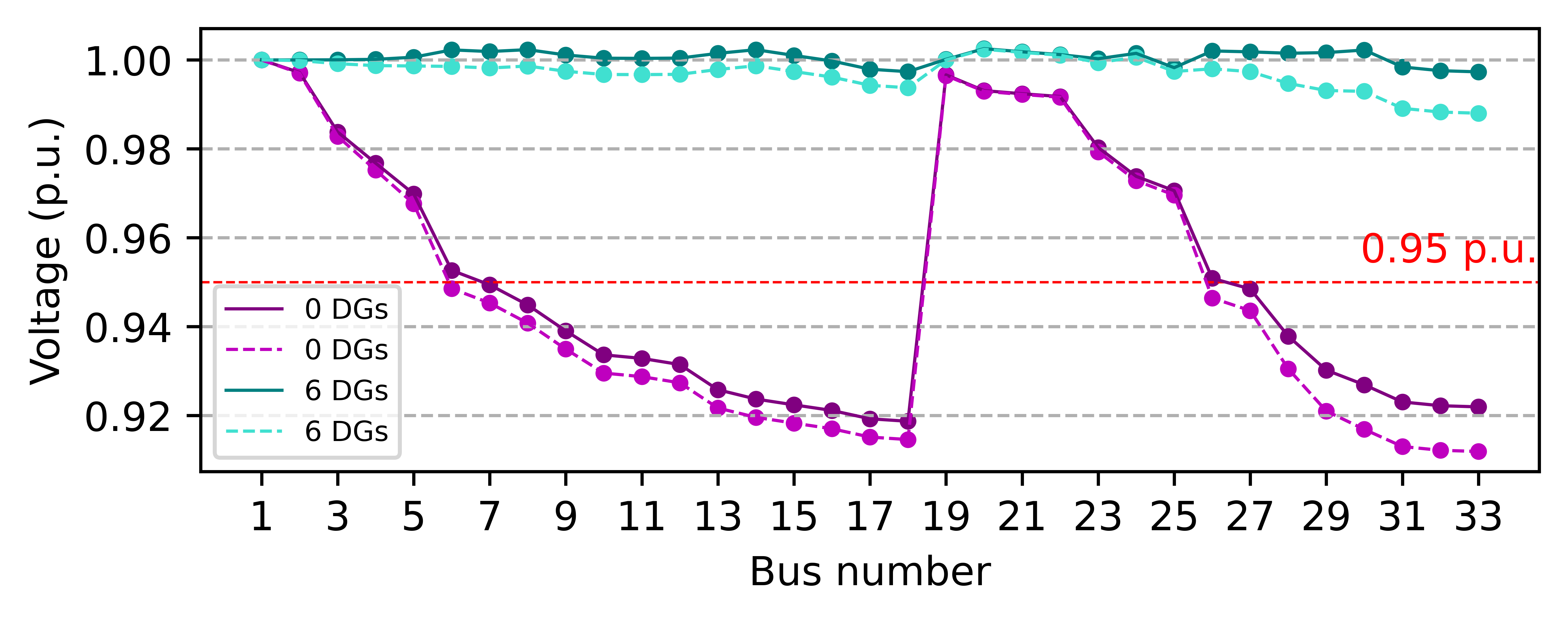}
    \vspace{-7mm}
    \caption{Bus voltages under a load increase at bus 31 of the IEEE 33-bus system. The solid lines and dash lines represent the voltage magnitudes before and after the load increase, respectively.}
    \vspace{-4mm}
    \label{fig:voltageLoad33}
\end{figure}

We further study the voltage variations due to the load increase for the buses in the systems which experience the highest variation of $\Delta L$, i.e., bus 8 and bus 31 in the 15-bus system and 33-bus system, respectively. In this experiment, we increase the load demand of only these two buses by 50\% and present the voltage profiles at each bus in Figs. \ref{fig:voltageLoad15} and \ref{fig:voltageLoad33}. It is clear that for both systems, the optimization contributed towards improved voltage levels. Despite small variations, all the voltage profiles are above $0.95$ p.u.

\vspace{-2mm}
\section{Conclusions}\label{s:conclusion}
\vspace{-0.5mm}
In this paper, we present a two-stage optimization strategy to reduce the power losses in the distribution grid. The approach first determines the optimal placement and active power outputs of the renewable-based DGs in the grid by solving a MISOCP optimization problem. In the second stage, we determine how DGs can provide the optimal reactive power according to different load demands. This latter stage achieves further power loss reduction. The proposed approach is tested on both IEEE 15-bus and IEEE 33-bus systems using DIgSILENT PowerFactory. The results verify that the power losses in the distribution grid can be significantly reduced. The voltage profiles of the distribution grid are also improved, and furthermore, the grid can support additional load due to losses savings. As future work, we plan to consider the load profile and weather-dependent variability of power outputs of DGs. Additionally, the application of the optimization framework in distribution systems with other grid typologies, such as meshed distribution grids, will be further explored.

\vspace{-1.5mm}

\bibliographystyle{IEEEtran}
\bibliography{refs}

\begin{thebibliography}{10}
\providecommand{\url}[1]{#1}
\csname url@samestyle\endcsname
\providecommand{\newblock}{\relax}
\providecommand{\bibinfo}[2]{#2}
\providecommand{\BIBentrySTDinterwordspacing}{\spaceskip=0pt\relax}
\providecommand{\BIBentryALTinterwordstretchfactor}{4}
\providecommand{\BIBentryALTinterwordspacing}{\spaceskip=\fontdimen2\font plus
\BIBentryALTinterwordstretchfactor\fontdimen3\font minus
  \fontdimen4\font\relax}
\providecommand{\BIBforeignlanguage}[2]{{%
\expandafter\ifx\csname l@#1\endcsname\relax
\typeout{** WARNING: IEEEtran.bst: No hyphenation pattern has been}%
\typeout{** loaded for the language `#1'. Using the pattern for}%
\typeout{** the default language instead.}%
\else
\language=\csname l@#1\endcsname
\fi
#2}}
\providecommand{\BIBdecl}{\relax}
\BIBdecl

\bibitem{world2015}
{World Bank Group}, ``Electric power transmission and distribution losses (\%
  of output),'' 2014, [Online]. Available:
  \url{https://data.worldbank.org/indicator/EG.ELC.LOSS.ZS}.

\bibitem{sultana2016review}
B.~Sultana \emph{et~al.}, ``Review on reliability improvement and power loss
  reduction in distribution system via network reconfiguration,''
  \emph{Renewable and sustainable energy reviews}, vol.~66, pp. 297--310, 2016.

\bibitem{NewYork2021}
{U.S. EIA}, ``New york electricity profile 2021,'' 2021, [Online]. Available:
  \url{https://www.eia.gov/electricity/state/newyork/}.

\bibitem{qiu1987new}
J.~Qiu and S.~Shahidehpour, ``A new approach for minimizing power losses and
  improving voltage profile,'' \emph{IEEE transactions on power systems},
  vol.~2, no.~2, pp. 287--295, 1987.

\bibitem{mamandur1981optimal}
K.~Mamandur and R.~Chenoweth, ``Optimal control of reactive power flow for
  improvements in voltage profiles and for real power loss minimization,''
  \emph{IEEE transactions on power apparatus and systems}, no.~7, pp.
  3185--3194, 1981.

\bibitem{badar2012reactive}
A.~Q. Badar, B.~Umre, and A.~Junghare, ``Reactive power control using dynamic
  particle swarm optimization for real power loss minimization,''
  \emph{International Journal of Electrical Power \& Energy Systems}, vol.~41,
  no.~1, pp. 133--136, 2012.

\bibitem{alkaabi2018short}
S.~S. Alkaabi, H.~H. Zeineldin, and V.~Khadkikar, ``Short-term reactive power
  planning to minimize cost of energy losses considering pv systems,''
  \emph{IEEE Transactions on Smart Grid}, vol.~10, no.~3, pp. 2923--2935, 2018.

\bibitem{deilami2011real}
S.~Deilami \emph{et~al.}, ``Real-time coordination of plug-in electric vehicle
  charging in smart grids to minimize power losses and improve voltage
  profile,'' \emph{IEEE Transactions on smart grid}, vol.~2, no.~3, pp.
  456--467, 2011.

\bibitem{luo2014real}
X.~Luo and K.~W. Chan, ``Real-time scheduling of electric vehicles charging in
  low-voltage residential distribution systems to minimise power losses and
  improve voltage profile,'' \emph{IET generation, transmission \&
  distribution}, vol.~8, no.~3, pp. 516--529, 2014.

\bibitem{singh2010influence}
M.~Singh, I.~Kar, and P.~Kumar, ``Influence of ev on grid power quality and
  optimizing the charging schedule to mitigate voltage imbalance and reduce
  power loss,'' in \emph{Proceedings of EPE-PEMC}.\hskip 1em plus 0.5em minus
  0.4em\relax IEEE, 2010.

\bibitem{baran1989network}
M.~E. Baran and F.~F. Wu, ``Network reconfiguration in distribution systems for
  loss reduction and load balancing,'' \emph{IEEE Transactions on Power
  delivery}, vol.~4, no.~2, pp. 1401--1407, 1989.

\bibitem{rao2012power}
R.~S. Rao \emph{et~al.}, ``Power loss minimization in distribution system using
  network reconfiguration in the presence of distributed generation,''
  \emph{IEEE transactions on power systems}, vol.~28, no.~1, pp. 317--325,
  2012.

\bibitem{pegado2019radial}
R.~Pegado \emph{et~al.}, ``Radial distribution network reconfiguration for
  power losses reduction based on improved selective bpso,'' \emph{Electric
  Power Systems Research}, vol. 169, pp. 206--213, 2019.

\bibitem{irena2023renewable}
IRENA(2023), ``Renewable capacity statistics 2023,'' International Renewable
  Energy Agency, Abu Dhabi, Tech. Rep., 2022.

\bibitem{lalitha2010optimal}
M.~P. Lalitha, V.~V. Reddy, and V.~Usha, ``Optimal dg placement for minimum
  real power loss in radial distribution systems using pso.'' \emph{Journal of
  Theoretical \& Applied Information Technology}, vol.~13, 2010.

\bibitem{kansal2013optimal}
S.~Kansal, V.~Kumar, and B.~Tyagi, ``Optimal placement of different type of dg
  sources in distribution networks,'' \emph{International Journal of Electrical
  Power \& Energy Systems}, vol.~53, pp. 752--760, 2013.

\bibitem{avar2021optimal}
A.~Avar and M.~K. Sheikh-El-Eslami, ``Optimal dg placement in power markets
  from dg owners’ perspective considering the impact of transmission costs,''
  \emph{Electric Power Systems Research}, vol. 196, p. 107218, 2021.

\bibitem{melgar2018adaptive}
O.~D. Melgar-Dominguez, M.~Pourakbari-Kasmaei, and J.~R.~S. Mantovani,
  ``Adaptive robust short-term planning of electrical distribution systems
  considering siting and sizing of renewable energy based dg units,''
  \emph{IEEE Transactions on Sustainable Energy}, vol.~10, no.~1, pp. 158--169,
  2018.

\bibitem{van2016linear}
T.~Van~Dao, S.~Chaitusaney, and H.~T.~N. Nguyen, ``Linear least-squares method
  for conservation voltage reduction in distribution systems with photovoltaic
  inverters,'' \emph{IEEE Transactions on Smart Grid}, vol.~8, no.~3, pp.
  1252--1263, 2016.

\bibitem{wang2013review}
Z.~Wang and J.~Wang, ``Review on implementation and assessment of conservation
  voltage reduction,'' \emph{IEEE Transactions on Power Systems}, vol.~29,
  no.~3, pp. 1306--1315, 2013.

\bibitem{ANSI1995ANSI}
ANSI, ``Ansi standard c84.1-1995 electric power systems and equipment voltage
  ratings (60 hz),'' 1995.

\bibitem{kennedy1991conservation}
B.~Kennedy and R.~Fletcher, ``Conservation voltage reduction (cvr) at snohomish
  county pud,'' \emph{IEEE Transactions on Power Systems}, vol.~6, no.~3, pp.
  986--998, 1991.

\bibitem{rahimi2012evaluation}
S.~Rahimi, M.~Marinelli, and F.~Silvestro, ``Evaluation of requirements for
  volt/var control and optimization function in distribution management
  systems,'' in \emph{IEEE ENERGYCON}.\hskip 1em plus 0.5em minus 0.4em\relax
  IEEE, 2012, pp. 331--336.

\bibitem{hossan2017comparison}
M.~S. Hossan, H.~M. Maruf, and B.~Chowdhury, ``Comparison of the zip load model
  and the exponential load model for cvr factor evaluation,'' in \emph{IEEE
  Power \& Energy Society General Meeting}, 2017, pp. 1--5.

\bibitem{nazir2020approximate}
F.~U. Nazir, B.~C. Pal, and R.~A. Jabr, ``Approximate load models for conic opf
  solvers,'' \emph{IEEE Transactions on Power Systems}, vol.~36, no.~1, pp.
  549--552, 2020.

\bibitem{sun2022multi}
X.~Sun \emph{et~al.}, ``A multi-mode data-driven volt/var control strategy with
  conservation voltage reduction in active distribution networks,'' \emph{IEEE
  Transactions on Sustainable Energy}, vol.~13, no.~2, pp. 1073--1085, 2022.

\bibitem{sudhakar2016modeling}
T.~Sudhakar \emph{et~al.}, ``Modeling and simulation of distribution network
  with the integration of distribution generator using matlab,'' \emph{Indian
  Journal of Science and Technology}, vol.~9, no.~12, pp. 1--7, 2016.

\bibitem{dolatabadi2020enhanced}
S.~H. Dolatabadi \emph{et~al.}, ``An enhanced ieee 33 bus benchmark test system
  for distribution system studies,'' \emph{IEEE Transactions on Power Systems},
  vol.~36, no.~3, pp. 2565--2572, 2020.

\end{thebibliography}
\end{document}